\documentclass[showpacs,nofootinbib]{revtex4}
\usepackage[latin1]{inputenc}
\usepackage[T1]{fontenc}
\usepackage{graphicx}
\usepackage{epsfig}
\usepackage{amsmath}
\usepackage{amssymb}
\usepackage{setspace}
\usepackage{color}
\usepackage{comment}
\usepackage{citesort}

\definecolor{grey}{gray}{0.95}

\def\bcommandlist #1 \ecommandlist{
\begin{table}[H]
 \fcolorbox{black}{grey}
	{\begin{tabular}{p{0.38\textwidth}p{0.58\textwidth}}
		#1
	\end{tabular}}
\end{table}}

\begin{document}

\newcommand{\mhalfo}{\frac{1}{2}}	%one half (half one)
\newcommand{\mhalf}[1]{\frac{#1}{2}}
\newcommand{\ka}{\kappa}
\newcommand{\equ}[1]{\begin{equation} #1 \end{equation}}
\newcommand{\eref}[1]{eq. (\ref{#1})}
\newcommand{\fref}[1]{Fig. \ref{#1}}
\newcommand{\ddotp}[1]{\frac{d^d #1}{(2\pi)^d}}	%d-dim. d over two pi
\newcommand{\nnnl}{\nonumber\\}	%nonumber new line
\newcommand{\G}[1]{\Gamma(#1)}
\newcommand{\nq}{\nu_1}	%q is under 1
\newcommand{\nw}{\nu_2}	%w is under 2
\newcommand{\nd}{\nu_3}	%d is under 3
\newcommand{\fig}[4]{\begin{figure}[#1]\centering\epsfig{file=#3}\caption{#4}\label{#2}\end{figure}}

\title{Algorithmic derivation of Dyson-Schwinger Equations}

\author{R. Alkofer\footnote{reinhard.alkofer@uni-graz.at}, M. Q. Huber\footnote{markus.huber@uni-graz.at}, K. Schwenzer\footnote{kai.schwenzer@uni-graz.at} \\}

% \ead{reinhard.alkofer@uni-graz.at}
% \author{Markus Q. Huber}
% \ead{markus.huber@uni-graz.at}
% \author{K. Schwenzer}
% \ead{kai.schwenzer@uni-graz.at}
\affiliation{Institut f\"ur Physik, Universit\"at Graz, Universit\"atsplatz 5, 8010 Graz, Austria}
% \date{\begin{large}\today\end{large}}
\date{\today}

\begin{abstract}
\noindent We present an algorithm for the derivation of Dyson-Schwinger equations of general theories that is suitable for an implementation within a symbolic programming language. Moreover, we introduce the \textit{Mathematica} package \textit{DoDSE}\footnote{The package can be obtained from physik.uni-graz.at/\textasciitilde mah/DoDSE.html.} which provides such an implementation. It derives the Dyson-Schwinger equations graphically once the interactions of the theory are specified. A few examples for the application of both the algorithm and the \textit{DoDSE} package are provided.
\end{abstract}

 % Field theory, quantum field theory

\maketitle

{\bf PROGRAM SUMMARY}

\begin{small}
\medskip 
\noindent
{\em Manuscript Title:} Algorithmic derivation of Dyson-Schwinger Equations\\
{\em Authors:} R. Alkofer, M. Q. Huber, K. Schwenzer                                                \\
{\em Program Title:} DoDSE                                          \\
% {\em Journal Reference:}                                      \\
  %Leave blank, supplied by Elsevier.
% {\em Catalogue identifier:}                                   \\
  %Leave blank, supplied by Elsevier.
{\em Licensing provisions:} none                                  \\
  %enter "none" if CPC non-profit use license is sufficient.
{\em Programming language:}  Mathematica 6 and higher                                  \\
{\em Computer:} all on which Mathematica is available                                              \\
  %Computer(s) for which program has been designed.
{\em Operating system:} all on which Mathematica is available                                       \\
  %Operating system(s) for which program has been designed.
% {\em RAM: } bytes                                              \\
  %RAM in bytes required to execute program with typical data.
% {\em Number of processors used:}                              \\
  %If more than one processor.
% {\em Supplementary material:}                                 \\
  % Fill in if necessary, otherwise leave out.
  % Please give some freely chosen keywords that we can use in a
  % cumulative keyword index.
  % see http://www.aip.org/pacs/pacs.html
{\em Classification: } 11.1  General, High Energy Physics and Computing \\
    \phantom{Classi}11.4 Quantum Electrodynamics\\
    \phantom{Classi}11.5 Quantum Chromodynamics, Lattice Gauge Theory\\
    \phantom{Classi}11.6 Phenomenological and Empirical Models and Theories                      \\
  %Classify using CPC Program Library Subject Index, see (
  % http://cpc.cs.qub.ac.uk/subjectIndex/SUBJECT_index.html)
  %e.g. 4.4 Feynman diagrams, 5 Computer Algebra.
% {\em External routines/libraries:}                                      \\
  % Fill in if necessary, otherwise leave out.
% {\em Subprograms used:}                                       \\
  %Fill in if necessary, otherwise leave out.
% {\em Catalogue identifier of previous version:}*              \\
  %Only required for a New Version summary, otherwise leave out.
% {\em Journal reference of previous version:}*                  \\
  %Only required for a New Version summary, otherwise leave out.
% {\em Does the new version supersede the previous version?:}*   \\
  %Only required for a New Version summary, otherwise leave out.
\noindent
{\em Nature of problem:} Derivation of Dyson-Schwinger equations for a theory with given interactions.\\
  %Describe the nature of the problem here.
%   \\
{\em Solution method:} Implementation of an algorithm for the derivation of Dyson-Schwinger equations.\\
  %Describe the method solution here.
%   \\
% {\em Reasons for the new version:}*\\
  %Only required for a New Version summary, otherwise leave out.
%    \\
% {\em Summary of revisions:}*\\
  %Only required for a New Version summary, otherwise leave out.
%    \\
% {\em Restrictions:}\\
  %Describe any restrictions on the complexity of the problem here.
%    \\
{\em Unusual features:} The results can be plotted as Feynman diagrams in Mathematica.\\
  %Describe any unusual features of the program/problem here.
 %  \\
% {\em Additional comments:}\\
  %Provide any additional comments here.
%    \\
{\em Running time:} Less than a second to minutes for Dyson-Schwinger equations of higher vertex functions.\\
  %Give an indication of the typical running time here.
   \\
% {\em References:}
% \begin{refnummer}
% \item Reference 1         % This is the reference list of the Program Summary
% \item Reference 2         % Type references in text as [1], [2], etc.
% \item Reference 3         % This list is different from the bibliography, which
                          % you can use in the Long Write-Up.
% \end{refnummer}
% * Items marked with an asterisk are only required for new versions
% of programs previously published in the CPC Program Library.\\
\end{small}

\section{Introduction}\label{sec:intro}

Correlation functions are the basic quantities in local quantum field theories and encode all physical information about the theory. They fulfill quantum equations of motion, conventionally called Dyson-Schwinger equations (DSEs) \cite{Dyson:1949ha,Schwinger:1951ex} which are related among each other and form a set of infinitely many coupled equations. Derived from the translational invariance of the path integral they are genuinely non-perturbative and describe the physics of the system on all scales.
%are not inherently restricted to certain momentum regions. 
This makes them a very useful tool for investigating aspects on which some alternative approaches fail. Perhaps the most prominent example is perturbation theory, which is not valid in the strong coupling regime. Since DSEs are likewise applicable in the weak coupling region they successfully extend the results of perturbation theory into the strong coupling domain.
An alternative non-perturbative tool, which can be used complementary to DSEs, are Monte-Carlo lattice simulations. Due to the discretization of space-time they have their limits for very low and very high momenta, the former being restricted by the size of the lattice and the latter by finite lattice spacings. DSEs, in contrast, are formulated in continuous spacetime and allow to study also the analytic structure and the infrared regime which is particularly important in an asymptotically free but confining gauge theory like quantum chromodynamics (QCD). 

However, DSEs also have their challenges. They represent strongly non-linear integral equations that are numerically involved. Moreover, as they form an infinite tower of equations, they have to be truncated. 
%This can be done by discarding certain vertex functions or making ans\"atze instead of using their %respective DSEs. 
Recently it turned out in the context of Landau gauge QCD that the leading order truncation based only on the propagator DSEs can miss important qualitative features that are encoded in the equations for the vertices. In particular, the quark-gluon vertex provides a novel mechanism for confinement and chiral symmetry breaking \cite{Alkofer:2008tt}, as well as anomalous mass generation \cite{Alkofer:2008et} in QCD. Yet, the DSEs for the vertices become increasingly complicated and correspondingly hard to obtain algebraically. Another complication is given by the necessity of gauge fixing and the additional degrees of freedom and interactions arising from the corresponding constraints. In particular, in non-covariant gauges, like Coulomb gauge or non-linear gauges like the maximally Abelian gauge, this increases the effort to derive the DSEs already at the propagator level considerably, see for example \cite{Watson:2006yq,VillalbaChavez:2008dv}, and calls for an algorithmic way to derive these fundamental equations. This is especially useful when working with actions that contain many different fields and interactions, as arise e.g. when symmetries are not manifestly realized 
%due to non-covariant gauge fixing
or in the case of necessary additional terms in the action. Examples for the latter are the four-ghost interaction required to ensure renormalizability in maximally Abelian gauge \cite{Min:1985bx,Fazio:2001rm} or generalized constructions of Lagrangians allowing additional terms as in ghost-antighost symmetric gauges \cite{Baulieu:1981sb,ThierryMieg:1985yv}.
A convenient way to derive DSEs also simplifies the comparison of different gauges required to obtain a more gauge independent picture of the basic underlying mechanisms. Finally, it is particularly useful in the context of an IR analysis where the IR scaling, which is important for long-range properties like confinement, can be abstracted from mere power counting.  

The aim of this paper is to present such an algorithmic derivation of Dyson-Schwinger equations. A similar aim has been extensively followed in perturbation theory where it resulted in the basically automatic computation of numerous physical processes to a given order, cf. e.g. \cite{Mertig:1990an,Hahn:2000kx,Caravaglios:1995cd,Kanaki:2000ey,Mangano:2002ea}. Here we partially extend this idea to the non-perturbative regime were such an automatic solution of the created equations is surely beyond our scope. Instead we present an algorithm for the derivation of the equations that is suited for implementation into a symbolic programming language. This algorithm is presented below in Sec. \ref{sec:algorithm} and is implemented in the \textit{Mathematica} package \textit{DoDSE}, which stands for {\em Derivation of Dyson-Schwinger Equations}. An example of how to use the algorithm is provided in Sec. \ref{sec:ghg-vertex}. Details on the \textit{DoDSE} package are presented in Sec. \ref{sec:DoDSE}.
%The aim of this paper is to provide an overview of the \textit{Mathematica} package \textit{DoDSE}, %which stands for {\em Derivation of Dyson-Schwinger Equations}. Therefore, we need an algorithm for %the derivation that is suited for implementation into a symbolic programming language. This algorithm %is presented below in Sec. \ref{sec:algorithm} and can of course also be used for deriving DSEs by %hand. 
Whereas a direct algebraic derivation can be quite a tedious task, with the symbolic and graphical notations employed, one can obtain DSEs for general actions with a relatively high number of interactions. Moreover, the presented algorithm operates directly on the level of the effective action and circumvents the tedious step to decompose connected into proper vertices necessary in a derivation on the level of the generating functional of full Green functions. 
We implemented this algorithm up to the diagrammatic level into \textit{DoDSE}. From the interactions of the theory, given in symbolic form, the code derives DSEs to the desired order. The outcome are symbolic representations of the Feynman diagrams encoding their topological structure and their symmetry factors. The last step, which has to be done manually, to get the full algebraic form of the DSEs is the replacement of the symbolic form by the explicit integral expression involving proper and bare correlation functions. For some applications of DSEs it is not even necessary to process the symbolic equations even further since they can be used directly, as is e.g. the case for scaling analyses. 
Finally, we note that the presented algorithm is in principle also applicable for the generation of a perturbative expansion by re-inserting the Dyson-Schwinger equations for dressed vertices and truncating at a given loop order.

\section{Deriving Dyson-Schwinger Equations in Symbolic Notation}
\label{sec:algorithm}

The algorithm presented here involves several abstractions. The first is the use of a symbolic notation that enables us to keep expressions relatively short. Writing out the equations in full detail, i.e. Lorentz and internal group indices as well as all coordinates in position space or momenta, leads to expressions that can easily hide the underlying basic structure. Secondly, we employ a super-field formalism that includes all irreducible fields of the theory into a single reducible multiplet. Finally, we exploit the fact that a local action can be expanded in the fields. Thereby the expansion coefficients, that can be operators in coordinate as well as internal space, can be left unspecified during the derivation and only have to be inserted in the end to obtain algebraic expressions for the graphical equations.
Thereby, the main steps can be done using diagrammatic replacement rules. All the techniques used in the following are not new, but combined they provide a powerful way of deriving DSEs conveniently and fast. As an explicit example we demonstrate the procedure for the two inequivalent DSEs of the ghost-gluon vertex in Landau gauge Yang-Mills theory in Sec. \ref{sec:ghg-vertex}. Before we explain the actual algorithm in Subsection \ref{ssec:algorithm} we make a short summary of used identities and employed conventions.

\subsection{Basic Identities and Conventions}

The basic object we will start from is the generating functional of one-particle irreducible (1PI) Green functions. In the following we do not specify any particular fields but will rather use a more compact notation with a super-field denoted by $\phi_i$. It represents a reducible multiplet containing all irreducible representations present in the action. We also introduce corresponding sources $J_{i}$. The multi-index $i$ includes the label of the particular irreducible representation, its internal indices and also the space-time dependence of the field. For example in Landau gauge QCD $\phi_i$ is the set of the gluon, ghost and quark fields: $\{A_\mu^a(x), c^a(x), \bar{c}^a(x), q^i(x), \bar{q}^i(x)\}$.  Where appropriate the Einstein convention for summation is understood and amended by integration.

The effective action is given by
\begin{align}
\Gamma[\Phi]&=-W[J]+\Phi_i J_i \; ,
\end{align}
where $W[J]$ is the generating functional of connected Green functions related to the functional path integral by
\begin{align}
Z[J]=\int D[\phi] e^{-S + \phi_j J_j}=e^{W[J]}
\end{align}
with $S$ being the gauge-fixed action.
The effective action $\Gamma[\Phi]$ depends on the averaged fields $\Phi$ in the presence of external currents $J$,
\begin{equation}
\Phi_{i}\equiv\left\langle \phi_{i}\right\rangle _{J}=\frac{\delta W}{\delta J_{i}}=Z[J]^{-1}\int D[\phi] \phi_i e^{-S + \phi_j J_j} \: .
\end{equation}
For Grassmann variables terms like $\phi_j J_j$ include the fields and sources in the usual order \footnote{Another possibility would be the use of a metric as for example used in ref. \cite{Pawlowski:2005xe}.}, i.e. for example in Landau gauge QCD we have $\phi_j J_j=\int dx (A_\mu^a(x) j_\mu^a(x) + \bar{\sigma}^a(x) c^a(x) + \bar{c}^a(x) \sigma^a(x) + \bar{\eta}^i(x) q^i(x) + \bar{q}^i(x) \eta^i(x))$.

For non-vanishing external sources (denoted by the index $J$) the propagator of the multiplet $\phi$ is
\begin{equation}
\Delta_{ij}^{J}\equiv\frac{\delta^{2}W}{\delta J_{i}\delta J_{j}}=\left(\frac{\delta^{2}\Gamma}{\delta\Phi_{i}\delta\Phi_{j}}\right)^{-1}\label{eq:super-prop}\end{equation}
but for vanishing external sources only some elements remain. For instance in Landau gauge QCD these would be a diagonal element for the gluon and off-diagonal elements for ghosts and quarks, but also mixed propagators can arise if allowed by the action. For the derivation of DSEs it is important to keep the general expressions and only at the end the external sources can be set to zero. Otherwise one would miss contributions.

Higher 1PI vertices are defined as derivatives of the effective action:
\begin{equation}
\Gamma_{i_{1}\cdots i_{n}}^{J}\equiv-\frac{\delta\Gamma}{\delta\Phi_{i_{1}}\cdots\delta\Phi_{i_{n}}} \; .
\end{equation}
In the following we will need the derivatives of fields, propagators and vertices with respect to fields. These are given by
\begin{subequations}\label{eq:derivatives}
\begin{align}
\frac{\delta}{\delta\Phi_{i}}\Phi_{j} & =\delta_{ij}\:,\\
\frac{\delta}{\delta\Phi_{i}}\Delta_{jk}^{J} & =\frac{\delta}{\delta\Phi_{i}}\left(\frac{\delta^{2}\Gamma}{\delta\Phi_{j}\delta\Phi_{k}}\right)^{-1}=-\left(\frac{\delta^{2}\Gamma}{\delta\Phi_{j}\delta\Phi_{m}}\right)^{-1}\left(\frac{\delta^{3}\Gamma}{\delta\Phi_{m}\delta\Phi_{i}\delta\Phi_{n}}\right)\left(\frac{\delta^{2}\Gamma}{\delta\Phi_{n}\delta\Phi_{k}}\right)^{-1}=\Delta_{jm}^{J}\Gamma_{min}^{J}\Delta_{nk}^{J}\:, \\
\frac{\delta}{\delta\Phi_{i}}\Gamma_{j_{1}\cdots j_{n}}^{J} & =-\frac{\delta\Gamma}{\delta\Phi_{i}\delta\Phi_{j_{1}}\cdots\delta\Phi_{j_{n}}}=\Gamma_{ij_{1}\cdots j_{n}}^{J}\:\end{align}
\end{subequations}
and are represented graphically in \fref{fig:diagRules}.

The physical correlation functions are obtained from the above expressions
when evaluated at the vacuum expectation values of the fields, corresponding
to vanishing external currents,
\begin{align*}
\Delta_{ij} & \equiv\left\langle \phi_{i}\phi_{j}\right\rangle =\left.\left(\frac{\delta^{2}\Gamma}{\delta\Phi_{i}\delta\Phi_{j}}\right)^{-1}\right|_{\Phi=\Phi^{0}}=\left.\Delta_{ij}^{J}\right|_{\Phi=\Phi^{0}}\:,\\
\Gamma_{i_{1}\cdots i_{n}} & \equiv\left\langle \phi_{i_{1}}\cdots\phi_{i_{n}}\right\rangle _{1PI}=-\left.\frac{\delta\Gamma}{\delta\Phi_{i_{1}}\cdots\delta\Phi_{i_{n}}}\right|_{\Phi=\Phi^{0}}=\left.\Gamma_{i_{1}\cdots i_{n}}^{J}\right|_{\Phi=\Phi^{0}}\:,
\end{align*}
where the vertices involve only proper, i.e. 1PI, diagrams. By construction the arising generating equations always involve one
bare vertex function
\begin{equation}
\label{eq:bare}
S_{i_{1}\cdots i_{n}}\equiv\left\langle \phi_{i_{1}}\cdots\phi_{i_{n}}\right\rangle _{1PI}^{0}=-\left.\frac{\delta S}{\delta\phi_{i_{1}}\cdots\delta\phi_{i_{n}}}\right|_{\Phi=\Phi^{0}}\quad,\quad\Phi_{i}^{0}\equiv\left.\frac{\delta W}{\delta J_{i}}\right|_{J=0}\:.
\end{equation}
We comment on the inclusion of fermions in Subsec. \ref{ssec:fermions}.
\begin{comment}
\begin{figure}
\begin{center}
\centering\epsfig{file=./DSE-diagRules,width=0.9\textwidth}
\end{center}
\caption{Diagrammatic rules for differentiating an external field, a propagator or a vertex. The circle with the cross denotes an external field, small blobs denote dressed propagators, and big blobs 1PI vertices. The double line represents the super-field $\phi$.}
\label{fig:diagRules}
\end{figure}
\end{comment}
\fig{t}{fig:diagRules}{./DSE-diagRules,width=0.9\textwidth}{Diagrammatic rules for differentiating an external field, a propagator or a vertex. The circle with the cross denotes an external field, small blobs denote dressed propagators, and big blobs 1PI vertices. The double line represents the super-field $\phi$.}%{0.9\textwidth}

\subsection{Algorithm}
\label{ssec:algorithm}

The functional DSEs \cite{Itzykson:1980ft,Roberts:1994dr,Alkofer:2000wg} are obtained by the invariance of the path integral with respect to variations of the integration variable $\phi$ \footnote{For a different approach using equal time commutation relations and Heisenberg's equation of motion see ref. \cite{Rivers:1988pi}.}:
\begin{align}\label{eq:DSE-Z}
\frac{\delta}{\delta \phi_i} Z[J]=&\int D[\phi] \left( -\frac{\delta S}{\delta \phi_i} + J_i \right) e^{-S + \phi_j J_j}=\nnnl
=&\left( -\frac{\delta S}{\delta \phi'_i}\Bigg\vert_{\phi'_i=\delta/\delta J_i} +J_i \right) Z[J]=0.
\end{align}
Substituting $Z[J]$ by $e^{W[J]}$ and using
\begin{align}
e^{-W[J]}\left(\frac{\delta}{\delta J_i}\right)e^{W[J]}= \frac{\delta W[J]}{\delta J_i}+\frac{\delta}{\delta J_i} 
\end{align}
after multiplication of \eref{eq:DSE-Z} from the left with $e^{-W[J]}$ we find
\begin{align}
-\frac{\delta S}{\delta \phi_i}\Bigg\vert_{\phi_i=\frac{\delta W[J]}{\delta J_i}+\frac{\delta}{\delta J_i}} +J_i=0.
\end{align}
This is the functional DSE for connected correlation functions. To get the corresponding version for 1PI functions we perform the Legendre transformation of $W$ with respect to all sources. Thereby $\delta W[J]/\delta J_i$ changes to $\Phi_i$ and $\delta/\delta J_i$ becomes
\begin{align}
\frac{\delta}{\delta J_i}=\frac{\delta \Phi_j}{\delta J_i} \frac{\delta}{\delta \Phi_j}=\frac{\delta}{\delta J_i} \frac{\delta W}{\delta J_j} \frac{\delta}{\delta \Phi_j}=\frac{\delta^2 W}{\delta J_i \delta J_j} \frac{\delta}{\delta \Phi_j}=
\Delta_{ij}^J \frac{\delta}{\delta \Phi_j}.
\end{align}
This yields
\begin{align}\label{eq:DSE-master}
-\frac{\delta S}{\delta \phi_i}\Bigg\vert_{\phi_i=\Phi_i+\Delta_{ij}^J  \, \delta/\delta \Phi_j} +\frac{\partial \Gamma}{\partial \Phi_i}=0.
\end{align}
We stress again that here the summation over the index $j$ includes summation over different fields.
%  We focus here only on DSEs for 1PI quantities, but one can derive them also for full or connected propagators and vertices \cite{Alkofer:2000wg,Rivers:1988pi}:
% \begin{align}\label{eq:DSE-master}
% &-\frac{\delta S}{\delta \phi'_i}\Bigg\vert_{\phi'_i=\phi_i+\tilde{\Delta}_{ij}\delta/\delta \phi_j}+\frac{\delta \Gamma}{\delta \phi_i}=0.
% \end{align}
To simplify the process of differentiating the action we expand it in the fields\footnote{We restrict ourselves here to quartic interactions since they are renormalizable in four dimensions, but there is no restriction on the order of the interactions, like e.g. five-point vertices, in the \textit{DoDSE} package.}. To stay as general as possible we again use the super-field $\phi$
\begin{align}\label{eq:S-exp}
S[\phi]=&\frac{1}{2!}S_{rs}\phi_r \phi_s - \frac{1}{3!} S_{rst} \phi_r \phi_s \phi_t -
	\frac{1}{4!}S_{rstu} \phi_r \phi_s \phi_t \phi_u.
\end{align}
but for explicit calculations one can alternatively use an expansion in the fields of the action at this point. 
The $S_{i\ldots}$ are the expansion coefficients and correspond to bare quantities (or the inverse ones in case of the propagators) of which for an explicit action usually several vanish. Their signs have been chosen in accordance with \eref{eq:bare}. For the propagators as well as for all derivative interactions these are operators in coordinate space and may have a non-trivial structure in the internal space as well. We now differentiate once with respect to a field $\phi_i$,
\equ{\label{eq:1st-deriv-DSE}
\frac{\delta{S}}{\delta{\phi_i}}=S_{is}\phi_s  - \frac{1}{2!} S_{ist} \phi_s \phi_t -
	\frac{1}{3!}S_{irst} \phi_s \phi_t \phi_u ,
}
and replace the field operators according to \eref{eq:DSE-master} by
\begin{align}\label{eq:replacement}
\phi_i \rightarrow \Phi_i + \Delta_{ij}^J \frac{\delta}{\delta \Phi_j}.
\end{align}
Performing the derivatives and employing \eref{eq:derivatives} we get the general {\em generating DSE} for 1PI Green functions, depicted in \fref{fig:functDSE}:
\fig{t}{fig:functDSE}{./1PI-DSE,width=0.7\textwidth}{The functional DSE for 1PI functions. Crosses in circles denote external fields. All internal propagators are 1PI and the big blob denotes a 1PI vertex function.}%{0.7\textwidth}
\begin{align}
\label{eq:genDSE}
\frac{\delta \Gamma}{\delta \Phi_i}=&S_{is} \Phi_s -  \mhalfo S_{ist}(\Phi_s \Phi_t + \Delta_{st}^J) + \nnnl
	&-\frac{1}{3!}S_{istu} (\Phi_s \Phi_t \Phi_u + 3 \Phi_s \Delta_{tu}^J + \Delta_{sv}^J \Delta_{tw}^J \Delta_{ux}^J \Gamma_{vwx}^J).
\end{align}
from which the DSEs of arbitrary Greens functions can be obtained. This equation can be decomposed into generating equations for the individual irreducible fields which by construction always involve one
bare vertex function. If we had calculated these equations by direct differentiation of the action as given in \eref{eq:DSE-master}, it would have been a tedious task to get the bare vertices and propagators in the final expression. In particular, in this case derivatives on $\delta$-functions occur that have to be properly resolved by partial integration. This is the advantage the formal expansion in fields entails. In this formalism everything is hidden in a single index of the fields and the details of the bare vertices only have to be specified in the end in order to obtain algebraic expressions for the DSEs. We have checked by an explicit computation in the case of non-Abelian gauge theory that both ways to compute \eref{eq:genDSE} indeed reproduce the same result.

The corresponding DSEs for arbitrary correlation functions are obtained by further functional derivatives of the generating equation \eref{eq:genDSE} which are computed via eqs. (\ref{eq:derivatives}).
From now on we can proceed by use of the diagrammatic rules given in \fref{fig:diagRules}, where all internal lines denote dressed propagators. 
%Note that in most diagrams we suppress the blobs of internal propagators, since they appear only 1PI %and never bare.
Here, a major advantage of the diagrammatic rules is that one does not have to take care of indices. 
%Furthermore the matter of directions of fermion lines is no reason to worry about until the end.
We would like to stress again that the appearing super-propagators have off-diagonal components corresponding to mixing of the irreducible fields. These mixed propagators are important in the derivation process, although they seem like an artificial complication. In "simple" cases as for example the three-gluon vertex DSE in Landau gauge the result would not change if we kept only "real" propagators. However, for the ghost-gluon vertex (see Subsec. \ref{sec:ghg-vertex} for details), for higher vertex functions and in certain gauges already for the propagators, some contributions would be missing.

The DSE for a generic two-point function is derived by performing another differentiation of the generating DSE \eref{eq:genDSE} using the diagrammatic replacement rules of \fref{fig:diagRules} in the corresponding diagrammatic representation \fref{fig:functDSE} in all possible ways. The result is shown in \fref{fig:2Point}. Proceeding to higher vertex functions the number of diagrams increases rapidly: For three-point vertices there are 15 generic diagrams and for four-point functions 60. For real applications it is therefore recommendable to exploit possible simplifications. First, the final number and form of graphs depend on the first differentiation in \eref{eq:genDSE} as the corresponding field determines which bare vertices appear in the diagrams. For example the DSE of the ghost-gluon vertex in Landau gauge QCD has only four terms, when the first derivative is performed with respect to a ghost field. In this case one can drop all diagrams with bare gluonic vertices. On the other hand, if one starts with the gluon field, all vertices have to be kept and one ends up with twelve graphs. Secondly, one can skip some diagrams taking into account where one is going. Simple examples are that for a three-point function we do not have to drag along the bare four-point vertex or diagrams with an external field can be dropped if no further derivatives with respect to this particular field follow.
\fig{t}{fig:2Point}{./genericDSE-2Point,width=0.95\textwidth}{The DSE for a generic two-point function.}%{0.95\textwidth}

\subsection{Inclusion of Grassmann Fields}
\label{ssec:fermions}

Anticommuting fields need slightly more care when performing derivatives to get the correct signs. In this subsection we denote Grassmann fields by $\psi$ and $\bar{\psi}$ with sources $\bar{\eta}$ and $\eta$, respectively. Let us first specify our convention for the derivatives with respect to Grassmann quantities. We choose left- and right-derivatives:
\begin{align}
& \frac{\delta}{\delta \psi} := \frac{\overset{\leftarrow}{\delta}}{\delta \psi}, & \frac{\delta}{\delta \bar{\psi}} := \frac{\overset{\rightarrow}{\delta}}{\delta \bar{\psi}}.
\end{align}
This entails that the proper definition of a Grassmann field propagator is
\begin{align}
\Delta^{\bar{\psi}\psi}_{ij}=\frac{\delta^2 W}{\delta {\bar{\eta}}_i \delta \eta_j}=\left( \frac{\delta^2 \Gamma}{\delta \bar{\psi}_i \delta \psi_j} \right)^{-1}.
\end{align}
In general derivatives always have to be ordered such that derivatives with respect to Grassmann fields are right of those with respect to anti-Grassmann fields. A quartic Grassmann interaction then has the form
\begin{align}
 \Gamma_{ijkl}=-\frac{\partial \Gamma}{\partial \bar{\psi}_i \partial \bar{\psi}_j \partial \psi_k \partial \psi_l}\Big|_{\bar{\psi}=\psi=0}.
\end{align}
For easier readability and also in correspondence with \textit{DoDSE} the indices of the $\Gamma$ do not reflect the order of how the derivatives are performed, but rather have the order in which the derivatives appear, i.e. that the differentiation with respect to $\bar{\psi}_j$ has to be performed before that with respect to $\bar{\psi}_i$, but $\psi_k$ comes before $\psi_l$.

When using the algorithm described above derivatives always act from the corresponding direction. Eqs. (\ref{eq:derivatives}) have to be amended by
\begin{subequations}\label{eq:derivsGrassmann}
\begin{align}
\label{eq:deriv-propagator-anti-Grassmann}\frac{\delta}{\delta\bar{\psi}_{i}}\Delta^J_{jk} & = \Delta^J_{jm}\Gamma^{J,\bar{\psi}\Phi\Phi}_{imn} \Delta^J_{nk},\\
\label{eq:deriv-propagator-Grassmann}\frac{\delta}{\delta\psi_{i}} \Delta^J_{jk} & = \Delta^J_{jm} \Gamma^{J,\Phi\Phi \psi}_{mni} \Delta^J_{nk},\\
\label{eq:deriv-vertex-anti-Grassmann}\frac{\delta}{\delta\bar{\psi}_{i}} \Gamma^J_{j_{1}\cdots j_{n}} & =\Gamma^J_{ij_{1}\cdots j_{n}},\\
\label{eq:deriv-vertex-Grassmann}\frac{\delta}{\delta \psi_{i}} \Gamma^J_{j_{1}\cdots j_{n}} & =\Gamma^J_{j_{1}\cdots j_{n}i},
\end{align}
\end{subequations}
where here the superscript $J$ denotes the dependence on all sources and the additional superscripts of the $\Gamma^J$ denote the fields corresponding to the indices.
This means in turn that at the end, when the external sources are set to zero and the reducible multiplet is decomposed into the irreducible, physical fields, some Grassmann derivatives can be unordered. Ordering them gives the signs expected normally for Feynman diagrams with fermion loops.
However, in some cases this algorithm is oversimplified when several super-fields are involved. Using diagrammatic rules this problem can be circumvented and the minus signs for closed fermion loops have to be added manually at the end. Since \textit{DoDSE} needs specific rules how to perform the derivatives, it may happen that the wrong sign appears for two-loop graphs.

A simple example for the change of sign is the quark respectively ghost loop in the gluon DSE of Landau gauge where $\psi=\{q,c\}$ . Replacing $\bar{\psi}_r$ by $\bar{\psi}_r+ \Delta^{J,\psi\bar{\psi}}_{rt}  \frac{\delta}{\delta \psi_t}$ in the corresponding part of the first derivative of the action with respect to $A_i$,
\begin{align}
-S^{A\bar{\psi}\psi}_{irs} \bar{\psi}_r \psi_s \rightarrow -S^{A\bar{\psi}\psi}_{irs} \left(\bar{\psi}_r \psi_s + \Delta^{J,\psi\bar{\psi}}_{rs} \right),
\end{align}
and differentiating once more with respect to $A_j$ yields
\begin{align}
% \frac{\delta^2 \Gamma}{\delta A_i \delta A_j}= - S^{A\bar{\psi}\psi}_{irs} \left(\frac{\delta \Gamma}{\delta \psi_r  \delta \bar{\psi}_{r'}}\right)^{-1} \left(\frac{\delta \Gamma}{\delta \psi_{s'} \delta \bar{\psi}_s}\right)^{-1} \frac{\delta \Gamma}{\delta A_i \delta \psi_{r'} \delta \bar{\psi}_{s'}}+\text{gluonic terms}.
\frac{\delta^2 \Gamma}{\delta A_i \delta A_j}= - S^{A\bar{\psi}\psi}_{irs} \Delta^{\psi\bar{\psi}}_{rr'} \Delta^{\psi\bar{\psi}}_{s's} \frac{\delta \Gamma}{\delta A_i \delta \psi_{r'} \delta \bar{\psi}_{s'}}+\text{gluonic terms}
\end{align}
Ordering the derivatives changes the sign of the expression and leads to the expected relative minus sign of closed fermion loops.
% However, the situation gets quite involved when several super-fields are involved. This may occasionally lead to wrong signs in DoDSE. As long as vertices with at most two Grassmann fields appear this procedure is unique. However, for quartic Grassmann interactions there is some arbitrariness in the definition of the order of the Grassmann fields among themselves and similar for the anti-fields. When using expressions directly this is no problem, since it reflects only the anti-commutativity of Grassmann fields, but it may seem awkward, if Feynman diagrams with "wrong" signs appear. This can be circumvented by proper definitions of the diagrammatic representations. \textit{DoDSE} does not perform such an ordering and may therefore give "unusual" signs in diagrams involving quartic or higher Grassmann interactions, but the underlying expressions are correct.

Finally we should comment on the expansion of the action when Grassmann fields are involved. First, we recommend to order Grassmann interactions such that all anti-Grassmann fields are left of the Grassmann fields. Second, the expansion coefficients are antisymmetric in the indices belonging to Grassmann fields. This entails that we can differentiate with respect to Grassmann fields as usual, e.g.
\begin{align}
\frac{\delta}{\delta \bar{\psi}_i}S^{\bar{\psi}\bar{\psi}\psi\psi}_{rstu} \bar{\psi}_r \bar{\psi}_s \psi_t \psi_u= S^{\bar{\psi}\bar{\psi}\psi\psi}_{istu} \bar{\psi}_s \psi_t \psi_u - S^{\bar{\psi}\bar{\psi}\psi\psi}_{ritu} \bar{\psi}_r \psi_t \psi_u= S^{\bar{\psi}\bar{\psi}\psi\psi}_{istu} \bar{\psi}_s \psi_t \psi_u + S^{\bar{\psi}\bar{\psi}\psi\psi}_{irtu} \bar{\psi}_r \psi_t \psi_u= 2 S^{\bar{\psi}\bar{\psi}\psi\psi}_{istu} \bar{\psi}_s \psi_t \psi_u.
\end{align}

\section{An example: The ghost-gluon vertex in Landau gauge Yang-Mills theory}
\label{sec:ghg-vertex}

For the derivation of the ghost-gluon vertex in Landau gauge \cite{Schleifenbaum:2004id} we start with \fref{fig:2Point}. As mentioned above, there are two DSEs for this vertex. These arise since the DSEs are derived from the invariance of the initial path integral under changes of the fields. The invariance for the individual fields in the theory yields different generating equations from \eref{eq:genDSE} that can by appropriate functional differentiation generate topologically distinct DSEs for the same mixed correlation function. Without approximations these equations should provide identical results but they may be affected differently by truncations. Moreover, the dynamics can be represented differently in the various equations as demonstrated for the quark-gluon vertex in \cite{Alkofer:2008tt}.
We start here with the simpler equation for the ghost-gluon vertex, i.~e. we take the derivative in \eref{eq:genDSE} with respect to the ghost field. 
%The derivation in full detail can be found in ref. \cite{Schleifenbaum:2004id}. 
Fig. \ref{fig:2Point} reduces in this case to\newline
% \begin{center}\includegraphics[width=0.37\textwidth]{./genericDSE-2Point-ghgh,width=0.37\textwidth}.\end{center}
\begin{center}\epsfig{file=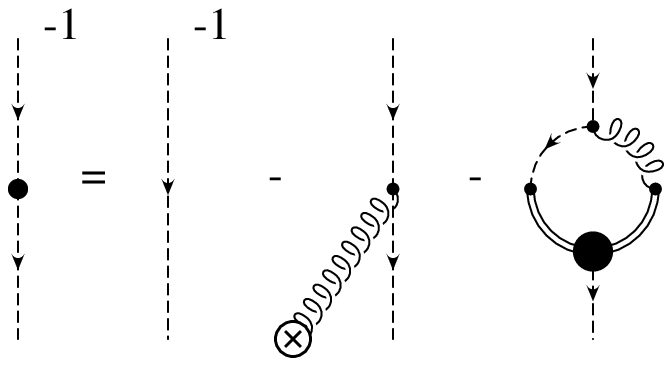,width=0.37\textwidth}.\end{center}
The symmetry factor $1/2$ vanished because there were two possibilities to plug the bare ghost-gluon vertex into the third diagram.
This is a nice example how the computation simplifies when the possible interactions are reduced by symmetries. In the present case the general diagrammatic rules specialize to\newline
% \begin{center}\includegraphics[width=0.95\textwidth]{./DSE-diagRules-gh.pdf}\end{center}
\begin{center}\epsfig{file=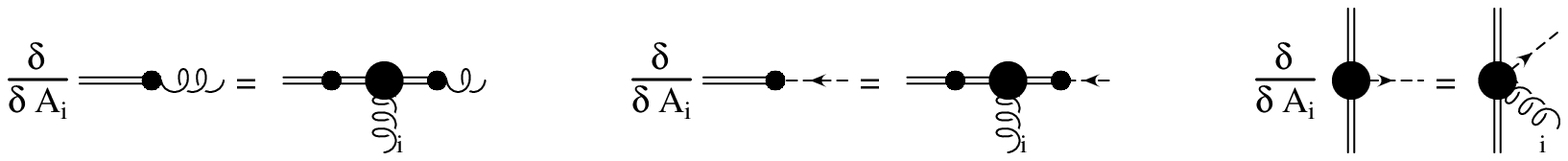,width=0.95\textwidth}\end{center}
which leads to the three-point expression
% \begin{center}\includegraphics[width=0.65\textwidth]{./1PI-ghg-DSE-superfields-gh.pdf}.\end{center}
\begin{center}\epsfig{file=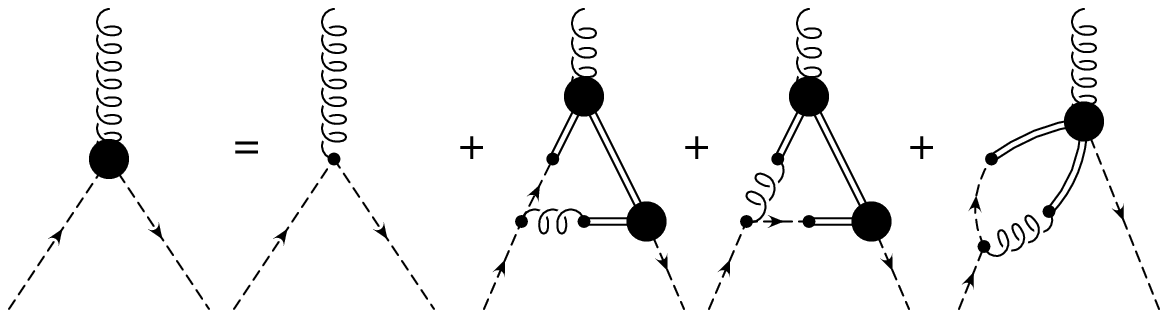,width=0.65\textwidth}.\end{center}
When we set the external sources to zero, the mixed propagators become irreducible gluon and ghost propagators. The pure super-field propagator in the second and third terms on the right-hand side yields a sum of different terms, when decomposed, but ghost number conservation allows here only vertices with the same number of ghost and anti-ghost legs. Therefore, for each diagram only one propagator can be realized. The final result is then the ghost-gluon vertex DSE:
% \begin{center}\includegraphics[width=0.65\textwidth]{./1PI-ghg-DSE-gh.pdf}.\end{center}
\begin{center}\epsfig{file=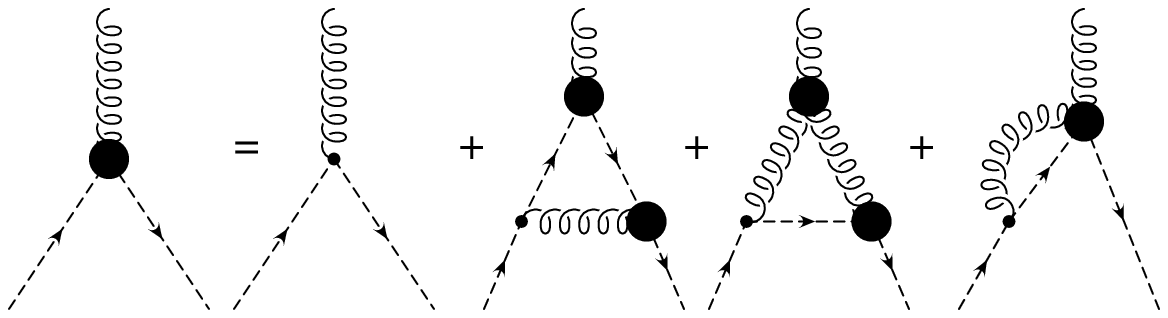,width=0.65\textwidth}.\end{center}

We obtain a second distinct version of the ghost-gluon DSE if we start differentiating with respect to the gluon field. In this case the super-field is important as will become evident below. The gluon is involved in all possible interactions of Landau gauge Yang-Mills theory, so \fref{fig:2Point} is not topologically simplified in this case.
%will not reduce to a simpler form as it did above. 
For brevity we skip diagrams that do not contribute to the ghost-gluon vertex (the tadpole and all incompatible tree graphs as well as the graph with the bare four-gluon vertex connected to an external field). The factor in front of the loop containing ghost-fields is changed from $1/2$ to $1$, because there are two possibilities to insert the bare ghost-gluon vertex as we have to consider the direction of fermion lines explicitly. The diagrams left are
% \begin{center}\includegraphics[width=0.65\textwidth]{./genericDSE-2Point-glgh.pdf}.\end{center}
\begin{center}\epsfig{file=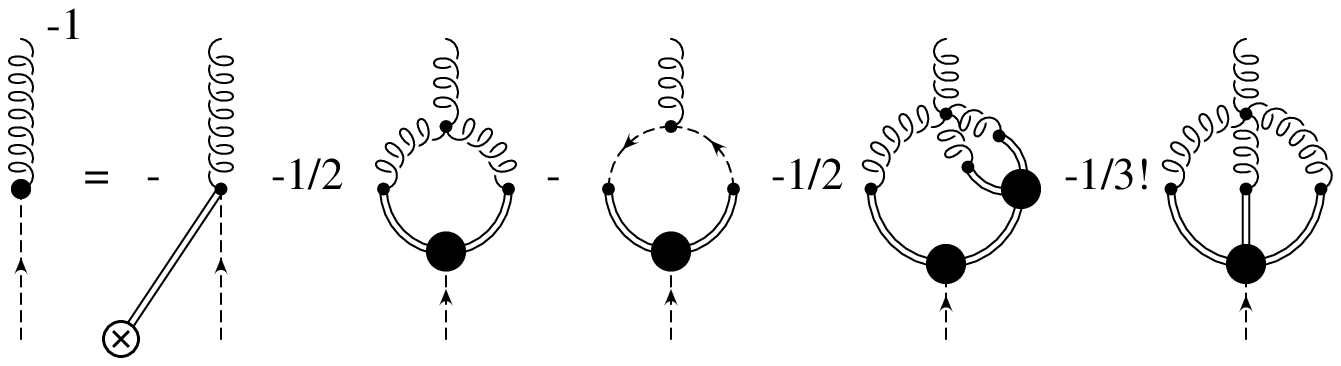,width=0.65\textwidth}.\end{center}
Differentiation with respect to the anti-ghost field yields
% \begin{center}\includegraphics[width=0.95\textwidth]{./1PI-ghg-DSE-superfields-gl.pdf}.\end{center}
\begin{center}\epsfig{file=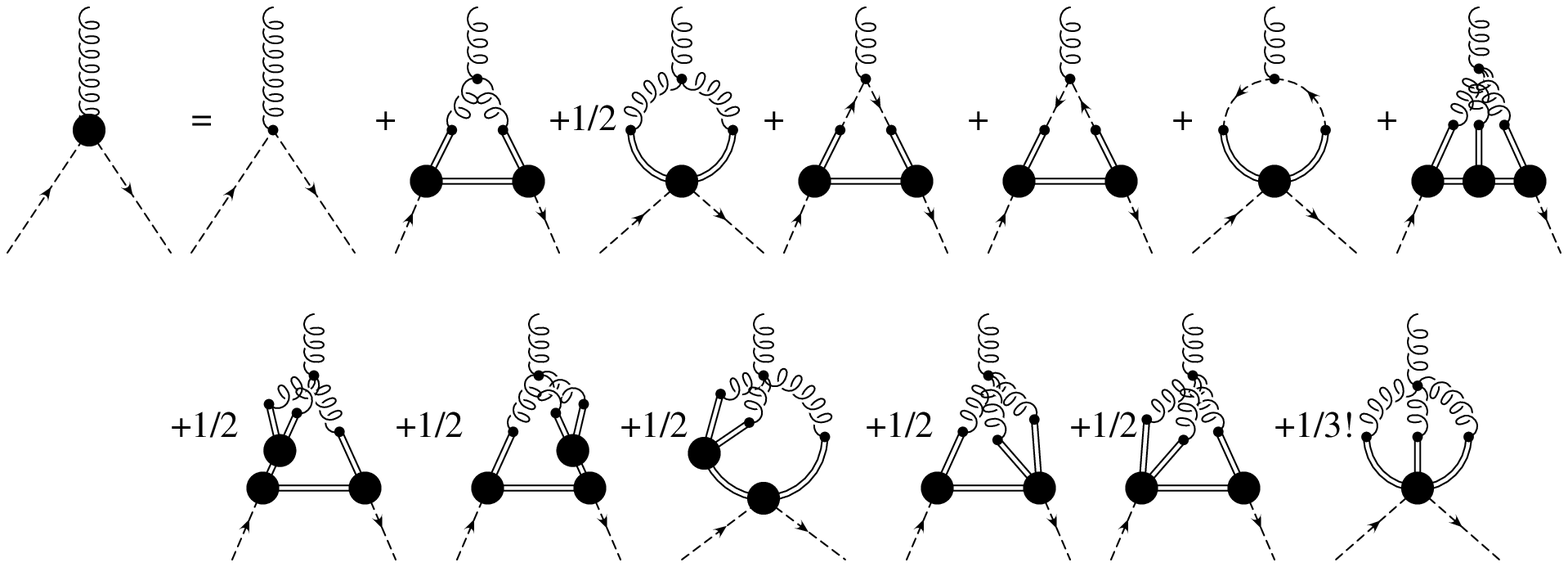,width=0.95\textwidth}.\end{center}
Again propagators partly involving the super-field are determined by the second field and pure super-field propagators by the symmetries of the vertices. Finally the second version of the ghost-gluon vertex DSE is obtained by setting the external sources to zero:
% \begin{center}\includegraphics[width=0.95\textwidth]{./1PI-ghg-DSE-gl.pdf}. \end{center}
\begin{center}\epsfig{file=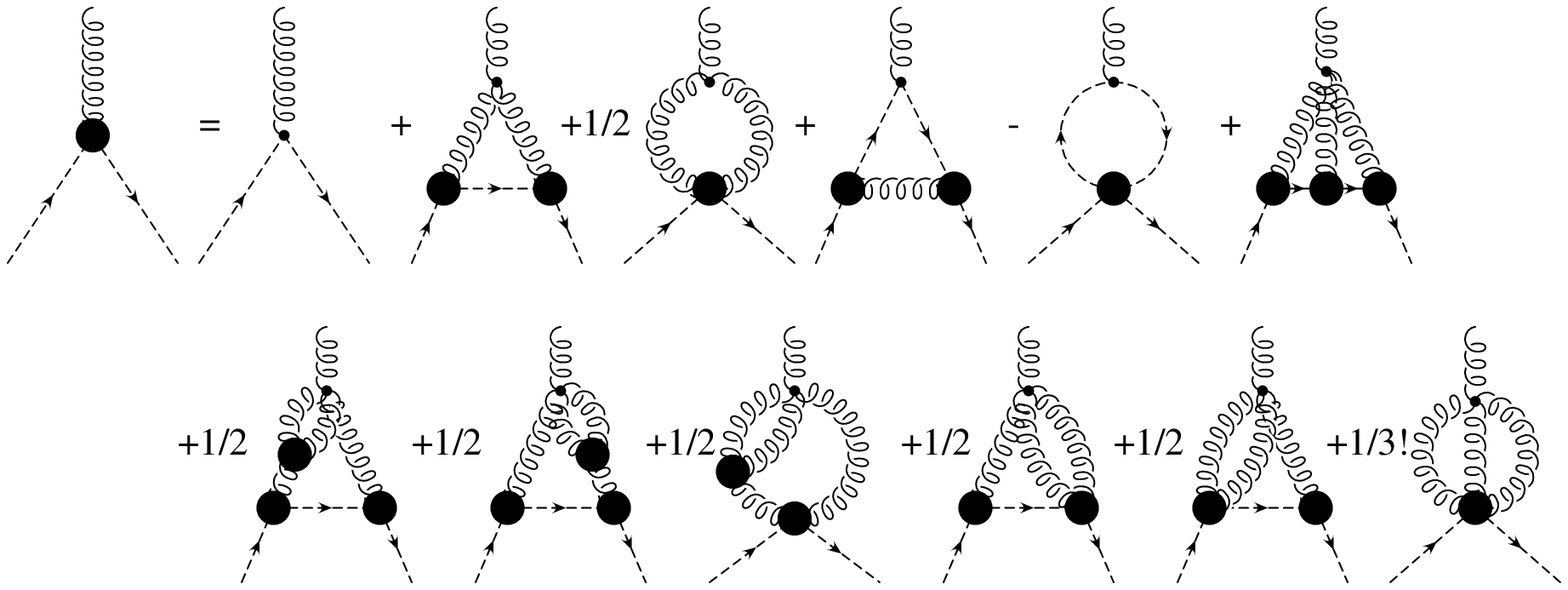,width=0.95\textwidth}. \end{center}

The derivation of the DSEs for the gluonic correlation functions can be done in a similar way. However, since the super-field cannot lead to internal ghost loops one can circumvent the super-field formalism and proceed with pure ghost and gluon propagators from the level of two-point functions on. For mixed Green functions this would omit some diagrams as was explicitly shown for the ghost-gluon vertex, where all diagrams with internal ghost lines - except the pure ghost loop (the fifth diagram on the right-hand side) - and the triangle diagrams would be missing.

\section{Derivation of Dyson-Schwinger Equations using Mathematica}
\label{sec:DoDSE}

As should have become clear the manual application of the algorithm becomes rather tedious for more complicated Green functions.
The algorithm explained in the Sec. \ref{sec:algorithm} is perfectly suitable for an implementation into a symbolic programming language like \textit{Mathematica} \cite{Wolfram:1999}. The functionality of the package \textit{DoDSE} is that the user enters the interactions of the theory and then all desired DSEs are automatically generated from it. In special cases it may in addition be necessary to provide further information about symmetries of vertices. Together with the package there is also a notebook (\textit{examples.nb}) available that contains the model used below as well as other examples like Landau gauge QCD.

\subsection{Technical Notes}

To load the package file \textit{DoDSE.m}, one either copies it to the subdirectory \textit{DoDSE} of \texttt{\$UserAddOns\-Directory}\footnote{Under Unix systems this is normally \textasciitilde/.Mathematica/Applications} and evaluates \verb|<< DoDSE`DoDSE`| (which would be the standard way of installation) or uses the \verb|Get| command to load it from any other place: \texttt{Get[pathToTheFile]}.
\textit{DoDSE} was developed under \textit{Mathematica} 6 %, release number 3 
and will not work with \textit{Mathematica}~5.2 or lower since some functions new to \textit{Mathematica}~6 are used. There exists no dedicated documentation within \textit{Mathematica}'s \textit{Documentation Center}, but help on single commands is available using the command \verb|?|, e.g. \verb|?DSEPlot|.

\subsection{Using \textit{DoDSE}}

For the derivation of DSEs via the \textit{DoDSE} package it is usually only necessary to specify the interactions of the given theory. This is done by a list whose elements are given by the individual propagators and vertices in the bare Lagrangian of the theory. These are in turn represented by lists containing the external fields of the corresponding correlation functions. 
\begin{comment}
necessary to define the following objects:
\begin{enumerate}
 \item The fields appearing in the action.
 \item An action.
 \item A test function to determine if a vertex respects the symmetries of the Lagrangian.
\end{enumerate}
\end{comment}
With this information the function \verb|doDSE| can now derive any DSE. 
The output can be used directly, put into a more convenient form using short notations for propagators and vertices or plotted with \verb|DSEPlot|. The complete calculation takes seconds, maybe minutes for vertex functions with more than four legs. As the package is written for \textit{Mathematica} we have a wide range of tools available for processing the results further, for instance the uniform IR scaling exponent of a graph (cf. e.g. \cite{Alkofer:2004it,Alkofer:2008jy}) can be derived directly from the output of \textit{DoDSE}.

%Having introduced the main objects of \textit{DoDSE}, we would like to explain 
In order to demonstrate the method to derive DSEs let us consider an example model.
Its action consists of four fields, two of them bosonic ($A$, $B$) and two fermionic ($c$, $d$):
\begin{align}
\mathcal{L}=&\mhalfo S^{AA}_{ij} A_i A_j+\mhalfo S^{BB}_{ij} B_i B_j+S^{\bar{c}c}_{ij}\bar{c}_i c_j+S^{\bar{d}d}_{ij}\bar{d}_i d_j+\nnnl
&-S^{A\bar{c}c}_{ijk} A_i \bar{c}_j c_k -\frac{1}{2}S^{AAB}_{ijk} A_i A_j B_k -\frac{1}{4}S^{AABB}_{ijkl} A_i A_j B_k B_l-\frac{1}{4!}S^{AAAA}_{ijkl} A_i A_j A_k A_l-S^{\bar{c}\bar{d}dc}_{ijkl} \bar{c}_i \bar{d}_j d_k c_l.
\end{align}
We wrote the quartic Grassmann interaction such that no additional minus sign occurs when differentiating with respect to $c$ and $\bar{c}$ or $d$ and $\bar{d}$.
The interactions of this action are entered in the form
\begin{verbatim}
ilist = {{A,A}, {B,B}, {cb,c}, {db,d}, {A,cb,c}, {A,A,B}, {A,A,B,B}, {A,A,A,A}, {cb,db,d,c}};
\end{verbatim}
\begin{comment}
It has to be entered into a notebook like this:
\begin{verbatim}
L = 1/2 op[ S[{A, i1}, {A, j1}], {A, i1}, {A, j1}] + 
    1/2 op[ S[{B, i1}, {B, j1}], {B, i1}, {B, j1}] + 
    + op[ S[{cb, i1}, {c, j1}], {cb, i1}, {c, j1}] + 
    + op[ S[{db, i1}, {d, j1}], {db, i1}, {d, j1}] + 
    + op[S[{A, i1}, {cb, j1}, {c, k1}], {A, i1}, {cb, j1}, {c, k1}] +
    + 1/(2 2) op[S[{A, i1}, {A, j1}, {B, k1}, {B, l1}],
        {A, i1}, {A, j1}, {B, k1}, {B, l1}] +
    + op[S[{cb, i1}, {db, j1}, {d, k1}, {c, l1}],
        {cb, i1}, {db, j1}, {d, k1}, {c, l1}] +
    + 1/4! op[S[{A, i1}, {A, j1}, {A, k1}, {A, l1}],
        {A, i1}, {A, j1}, {A, k1}, {A, l1}];
\end{verbatim}
\end{comment}
With this representation of the theory, we can start deriving the propagator DSEs using the function \verb|doDSE| which generates the non-trivial right hand side of the DSE for the corresponding correlation function. As arguments it takes the list of interactions \verb|ilist|, and a list of the fields with their respective indices included in the correlator for which we want to derive the DSE. In general, the order of the elements in the list of the correlation function for which the DSE is derived determines the order in which the individual functional derivatives are taken. When different fields are involved this can result in distinct DSEs, as was the case for the ghost-gluon vertex discussed in the last section. The corresponding commands for the propagators read:
\begin{verbatim}
AADSE = doDSE[ilist, {A, A}];
BBDSE = doDSE[ilist, {B, B}];
ccDSE = doDSE[ilist, {{c, i}, {cb, j}}];
ddDSE = doDSE[ilist, {{d, i}, {db, j}}];
\end{verbatim}
It is not necessary to give the indices of the fields, but it can be done as shown above for the two fermion DSEs. Alternatively one can derive several DSEs by passing a list to \verb|doDSE|, i.e. the following line derives one DSE for all primitively divergent vertex functions:
\begin{verbatim}
DSEs = doDSE[ilist, ilist];
\end{verbatim}

We can bring the rather long output into a more readable form with \texttt{shortExpression}:
\begin{verbatim}
shortExpression[AADSE]
\end{verbatim}
or equivalently
\begin{verbatim}
sE[AADSE]
\end{verbatim}
\begin{align}
&S_{\text{i j}}^{\text{A A}}-\frac{1}{2} \left(S_{\text{i j r1 s1}}^{\text{A A A A}} \Delta _{\text{r1 s1}}^{\text{A A}}\right)-\frac{1}{2} \left(S_{\text{i j r1 s1}}^{\text{A A B B}} \Delta _{\text{r1 s1}}^{\text{B B}}\right)-S_{\text{i r1 s1}}^{\text{A A B}} \Gamma _{\text{j t1 u1}}^{\text{A A B}} \Delta _{\text{r1 t1}}^{\text{A A}} \Delta _{\text{s1 u1}}^{\text{B B}}+S_{\text{i r1 s1}}^{\text{A cb c}} \Gamma _{\text{j t1 u1}}^{\text{A cb c}} \Delta _{\text{s1 t1}}^{\text{c cb}} \Delta _{\text{u1 r1}}^{\text{c cb}} \nonumber \\
&-\frac{1}{6} \left(S_{\text{i r1 r2 s1}}^{\text{A A A A}} \Gamma _{\text{j s2 t2 u2}}^{\text{A A A A}} \Delta _{\text{r1 s2}}^{\text{A A}} \Delta _{\text{r2 t2}}^{\text{A A}} \Delta _{\text{s1 u2}}^{\text{A A}}\right)-\frac{1}{2} \left(S_{\text{i r1 r2 s1}}^{\text{A A B B}} \Gamma _{\text{j s2 t2 u2}}^{\text{A A B B}} \Delta _{\text{r1 s2}}^{\text{A A}} \Delta _{\text{r2 t2}}^{\text{B B}} \Delta _{\text{s1 u2}}^{\text{B B}}\right)\nonumber\\
&-S_{\text{i r1 r2 s1}}^{\text{A A B B}} \Gamma _{\text{j u2 v1}}^{\text{A B A}} \Gamma _{\text{s2 t2 u1}}^{\text{A A B}} \Delta _{\text{r1 s2}}^{\text{A A}} \Delta _{\text{r2 u2}}^{\text{B B}} \Delta _{\text{s1 u1}}^{\text{B B}} \Delta _{\text{t2 v1}}^{\text{A A}}-S_{\text{i r1 r2 s1}}^{\text{A A B B}} \Gamma _{\text{j u2 v1}}^{\text{A B B}} \Gamma _{\text{s2 t2 u1}}^{\text{A B B}} \Delta _{\text{r1 s2}}^{\text{A A}} \Delta _{\text{r2 u2}}^{\text{B B}} \Delta _{\text{s1 u1}}^{\text{B B}} \Delta _{\text{t2 v1}}^{\text{B B}}\nonumber\\
&-\frac{1}{2} \left(S_{\text{i r1 r2 s1}}^{\text{A A A A}} \Gamma _{\text{j s2 t1}}^{\text{A A A}} \Gamma _{\text{u1 v2 w1}}^{\text{A A A}} \Delta _{\text{r1 s2}}^{\text{A A}} \Delta _{\text{r2 v2}}^{\text{A A}} \Delta _{\text{s1 w1}}^{\text{A A}} \Delta _{\text{u1 t1}}^{\text{A A}}\right)-\frac{1}{2} \left(S_{\text{i r1 r2 s1}}^{\text{A A B B}} \Gamma _{\text{j s2 t1}}^{\text{A A A}} \Gamma _{\text{u1 v2 w1}}^{\text{A B B}} \Delta _{\text{r1 s2}}^{\text{A A}} \Delta _{\text{r2 v2}}^{\text{B B}} \Delta _{\text{s1 w1}}^{\text{B B}} \Delta _{\text{u1 t1}}^{\text{A A}}\right) \nonumber\\
&-\frac{1}{2} \left(S_{\text{i r1 r2 s1}}^{\text{A A A A}} \Gamma _{\text{j s2 t1}}^{\text{A A B}} \Gamma _{\text{u1 v2 w1}}^{\text{B A A}} \Delta _{\text{r1 s2}}^{\text{A A}} \Delta _{\text{r2 v2}}^{\text{A A}} \Delta _{\text{s1 w1}}^{\text{A A}} \Delta _{\text{u1 t1}}^{\text{B B}}\right)-\frac{1}{2} \left(S_{\text{i r1 r2 s1}}^{\text{A A B B}} \Gamma _{\text{j s2 t1}}^{\text{A A B}} \Gamma _{\text{u1 v2 w1}}^{\text{B B B}} \Delta _{\text{r1 s2}}^{\text{A A}} \Delta _{\text{r2 v2}}^{\text{B B}} \Delta _{\text{s1 w1}}^{\text{B B}} \Delta _{\text{u1 t1}}^{\text{B B}}\right)\end{align}
\verb|shortExpression| uses the \textit{Mathematica} function \verb|Style| and accepts its corresponding options, e.~g. colors or \verb|FontSize|. The symbols used for propagators and vertices are set with the variables \verb|$bareVertexSymbol|, \verb|$vertexSymbol| and \verb|$propagatorSymbol|. The standard settings are $S$, $\Gamma$ and $\Delta$. The subscripts of these expressions are the indices, whereas the corresponding fields can be found in the superscript.
Alternatively one can plot the DSEs with \verb|DSEPlot|. To improve the representation it is advantageous to define a few graphics primitives for each field in a list of the form \texttt{\{\{field1, primitives1\}, \{field2, primitives2\}, ...\}}:
\begin{verbatim}
fieldRules = {{A, Red}, {B, Green, Dashed}, {c, Blue, Dotted},
  {d, Orange, Dashing[{0.02, 0.01}]}};
\end{verbatim}
These primitives can be used in \verb|DSEPlot|:
\begin{verbatim}
DSEPlot[AADSE, ilist, fieldRules]
\end{verbatim}
% \begin{center}\includegraphics[width=\textwidth]{./AADSE.pdf}\end{center}
\begin{center}\epsfig{file=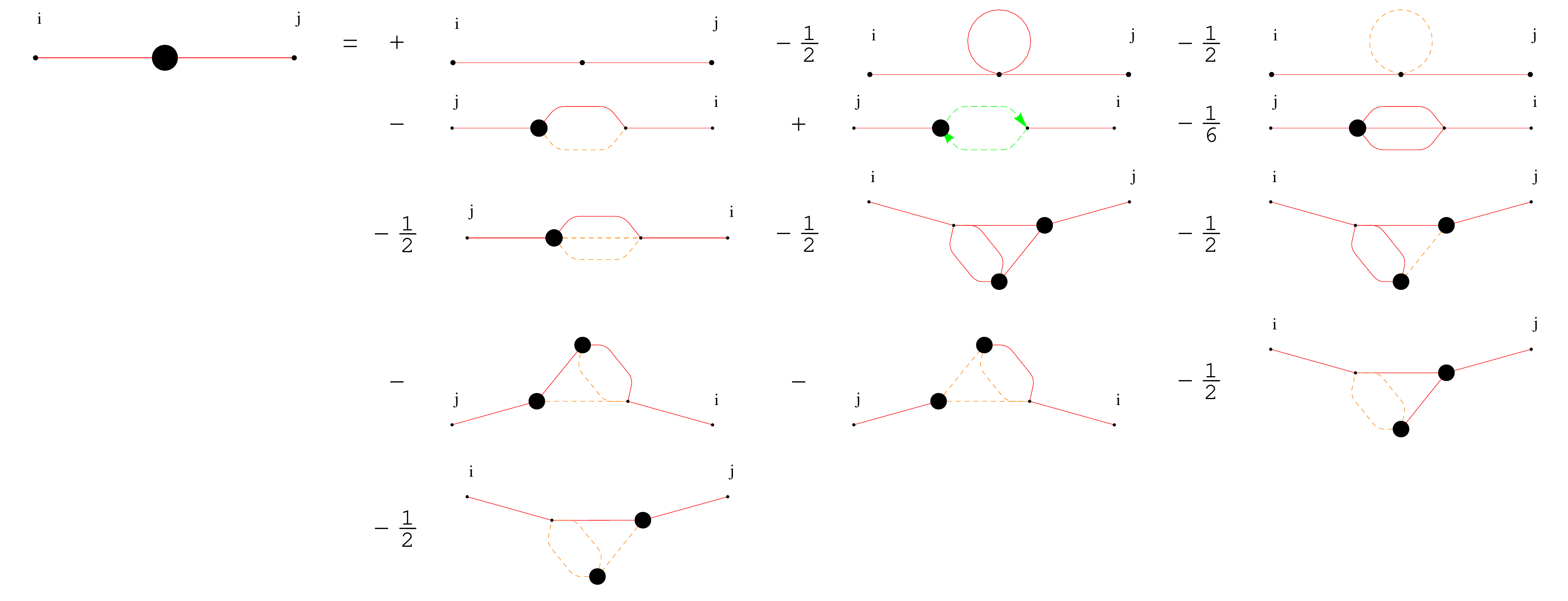,width=\textwidth}\end{center}
Since \verb|DSEPlot| uses the \textit{Mathematica} function \verb|GraphPlot| it inherits its corresponding options which are unfortunately quite limited.
%options for changing the output are quite limited. 
For example \verb|GraphPlot| may show the external points at different places for different graphs. It is possible to provide specific coordinates for the external points, but then some graphs may appear with overlapping internal lines. For many internal lines it can also happen that \verb|GraphPlot| draws lines above each other. For this reason \verb|DSEPlot| should be considered more a tool for showing the results of \textit{DoDSE} than a means of creating nice Feynman diagrams. However, the latter is also possible to a limited extend using the available options. In the generated diagrams all propagators are dressed. 1PI vertices are denoted by a large blob, whereas bare vertices are drawn without a blob. Fermionic fields have an arrow denoting their direction. One can invoke \verb|DSEPlot| also without the additional graphics primitives:
\begin{verbatim}
DSEPlot[AADSE, ilist]
\end{verbatim}
As a result text labels are attached to the individual propagators denoting their type.

Alternatively one can get a list of all graphs by setting the option \verb|output| to \verb|List|. Other options are \verb|indexStyle| and \verb|factorStyle|, which determine the styles of the indices and prefactors.

\subsection{Implementation}

In the following we will describe the individual operation steps of the function \verb|doDSE| for the interested reader who wants to do a step-by-step calculation for example to check own calculations.
Furthermore using the individual routines one has several additional possibilities on which we comment below. We also explain the representation of integrals in \textit{DoDSE}.

The function \verb|doDSE| performs the following operations:
\begin{enumerate}
 \item It converts the list of interactions to the internal representation of the action with \verb|generateAction|. Alternatively one can directly give such an expression as argument to \verb|doDSE|.
 \item The first derivative is performed directly on the action using \verb|deriv|, see \eref{eq:DSE-master}.
 \item Then the replacement according to \eref{eq:replacement} is done with \verb|replacementCalc|.
 \item At this point some graphs may appear several times. \verb|identifyGraph| adds them up to avoid redundant calculations and get the symmetry factors right.
 \item The necessary number of further derivatives can be worked out with \verb|deriv|.
 \item The external sources are set to zero using \verb|setSourcesZero|.
 \item For Grassmann fields it may be necessary to order the derivatives with \verb|orderFermions| such that anti-fields are left from fields thereby possibly changing the sign.
 \item Due to our definition of the vertex functions the left hand side contains a minus sign so that the final expression is multiplied by $-1$ for all vertex DSEs.
\end{enumerate}

In the package several objects are introduced for representing Feynman diagrams. The smallest units are fields and indices. Typically they are combined in a list like \verb|{A,i}|, where \verb|A| is a field and \verb|i| its index. They can stand for external fields or internal ones. These lists can be grouped to form propagators and vertices, denoted by \verb|P| and \verb|V| (1PI) or \verb|S| (bare). Finally external fields, propagators and vertices are combined in an object called \verb|op|, the biggest object, which represents individual diagrams. In it indices occurring twice are summed and integrated over. Note that a field is external if it is an argument of \verb|op|, while fields in vertices represent their legs. An example for a graph containing a propagator and an external field \verb|{A,j}| is
\begin{verbatim}
op[S[{A,i},{A,j},{A,k},{A,l}], P[{A,k},{A,l}], {A,j}].
\end{verbatim}
The index \verb|i| only appears once and is therefore an external leg.
The \verb|op|-function is the main object used internally for the computations and is also returned in the output of \verb|doDSE|. It can be used like one would expect from something representing a graph, for instance one can sum up several of them or it vanishes if one of its arguments is zero. Furthermore it splits up if one argument is a sum. For Grassmann fields there is a peculiarity in the notation: To make it as easy as possible to read \verb|op|-expressions, when two indices appear that are summed over also the corresponding fields are written identical, although in the notation above there should be the anti-field in the propagator for Grassmann fields. This means that the standard Grassmann propagator is defined as
\begin{verbatim}
P[{c,i}, {cb,j}],
\end{verbatim}
where \verb|c| is a Grassmann field and \verb|cb| the corresponding anti-field. Vertices are defined as expected, i.e. anti-field derivatives are left from field derivatives. As inverse bare propagators are denoted by \verb|S|, the function of bare vertices, they have the convention of vertices in contrast to dressed propagators.

\subsection{Advanced Options}

In some cases the features of \textit{DoDSE} described so far are not sufficient. For instance the occurrence of mixed propagators or symmetries not apparent in the generic expansion of the fields pose more intricate problems. This happens for example in the maximal Abelian gauge, where three-point functions involving three diagonal gluons are not allowed due to the color algebra but by default they are constructed by \textit{DoDSE}, which does not know about the color structure. These vertices have to be removed by appropriate restrictions.

To understand the necessity of vertex test functions and explicit definition of propagators as explained below, it may be helpful to know about the use of the super-field in \textit{DoDSE}. During the derivation of a DSE the super-field\footnote{Its standard name is $\phi$, but that can be changed by redefining the variable \texttt{\$dummyField}. As long as one does not do step-by-step calculations it should never appear in output.} can occur in two places: vertices and propagators. For the latter a list of rules of possible replacements is needed. Without further input \textit{DoDSE} uses only those propagators that appear in the Lagrangian. With these rules the fields in the propagators and vertices are replaced. However, as some of the vertices are forbidden by the symmetries of the action, further tests, called vertex test functions, are needed to assess if a vertex can exist in the theory.

In general \textit{DoDSE} performs two standard tests. The first checks for number conservation of individual Grassmann field species. It can be disabled with the option \verb|doGrassmannTest->False| for example to allow mixing. The second one makes sure that bosons with a discrete symmetry $\phi \leftrightarrow -\phi$ in the Lagrangian only appear in vertices that respect that symmetry. These standard tests can be amended by user-defined test functions, which take a vertex as argument and give \verb|True| or \verb|False|. To employ such functions they are given as argument to \verb|doDSE| (or \verb|setSourcesZero|). Vertex test functions allow also to truncate the system of DSEs, e.g. by forbidding vertices with a certain number of legs. We demonstrate this with a truncation of the AA DSE from above by defining a test function that only allows three-point vertices:
\begin{verbatim}
vertexTest[a_V] := Length@a==3
\end{verbatim}
Called with this restriction
\begin{verbatim}
AADSETruncated = doDSE [ilist, {A, A}, vertexTest];
\end{verbatim}
the number of diagrams reduces from $13$ to $11$, because the two sunset diagrams vanished.

In the case of mixed propagators in the Lagrangian it is necessary to provide the additional option \verb|specificFieldDefinitions| to \verb|doDSE|, e.g.
\begin{verbatim}
AADSEMixedL = doDSE[{{A, A}, {A, B}, {B, B}, {cb, c}, {A, cb, c}, {A, A, B}, {A, A, A}},
 {A, A}, specificFieldDefinitions->{A, B, {c, cb}}];
\end{verbatim}
If this option was not given, \verb|doDSE| would assume that \verb|A| and \verb|B| are fermion and anti-fermion. The list \verb|{A, B, {c, cb}}| tells \verb|doDSE| that \verb|A| and \verb|B| are bosons, \verb|c| is a fermion and \verb|cb| the corresponding anti-fermion.

If the theory allows mixed propagators not present in the Lagrangian, one has to provide a list of all possible propagators to \verb|doDSE|, e.g.
\begin{verbatim}
AADSEmixed = doDSE[ilist, {A, A},  {{A,A}, {A,B}, {B,B}, {c,cb}, {d,db}}];
\end{verbatim}
where we allowed a propagator between the \verb|A| and \verb|B| field.
The number of terms in the generated DSEs increases correspondingly.
The presence of mixed propagators has also a drawback: The final result may contain some terms several times. To add them up one employs the function \verb|identifyGraphs| with an additional option:
\begin{verbatim}
AADSEMixedLId = identifyGraphs[AADSEMixedL, compareFunction -> compareGraphs2];
\end{verbatim}
Since \verb|compareGraphs2| can take quite long, \verb|doDSE| normally uses \verb|compareGraphs|, which is not adequate for the situation with mixed propagators. The number of terms reduces as can be checked with
\begin{verbatim}
countTerms/@{AADSEMixedL, AADSEMixedLId} 
\end{verbatim}

In principle it is possible to derive a DSE step by step instead of using \verb|doDSE|. Thereby one can track every single step of the calculation corresponding to the algorithm described above and perform further manipulations, e.g. keeping some external fields. We refrain here from an explanation how to do this and refer the interested reader to the notebook \textit{examples.nb}, which is available together with the package. Finally there are a few tools available within \textit{DoDSE} that allow to check the syntax of expressions or to obtain information about the fields. They are listed in the appendix.

\section{Summary}

In this article we presented an algorithm to derive Dyson-Schwinger equations in a convenient way, in which one does not have to deal with the usual abundance of indices and integrals. This is achieved by graphical rules for performing derivatives that allow a quick and straight forward derivation even of higher vertex functions. We used this algorithm in the \textit{Mathematica} package \textit{DoDSE} that can give DSEs once the interactions of the theory are specified. It proves especially useful for theories with many interactions or higher vertex functions in general as the number of terms grows considerably. This should in particular help to analyze gauge theories in different gauges in order to obtain a more gauge independent picture of the described physics.

\section*{Acknowledgments}
It is a pleasure to thank Axel~Maas and Selym~Villalba-Chavez for valuable discussions.
M.~Q.~H. is supported by the  Doktoratskolleg 
``Hadrons in Vacuum, Nuclei and Stars'' of the Austrian science fund 
(FWF) under contract W1203-N08.
K.~S. acknowledges support from the FWF under contract M979-N16, and 
R.~A. from the German research foundation (DFG) under contract AL 279/5-2.

\appendix

\section*{Appendix: Tables of Functions}
\label{sec:tables}

In the following we provide lists of all public functions of \textit{DoDSE}. We give their syntax and a short explanation what they do. When the package is loaded one can always get help on the commands and their syntax within a notebook using the command \verb|?|, e.g. \verb|?doDSE|.

\bcommandlist
\multicolumn{2}{c}{\textbf{Main functions}}\\
 \hline
  Command & Description\\
\hline
\texttt{doDSE[ilist, clist]} & Derives the DSE for the correlation function \texttt{clist} for a theory with interactions \texttt{ilist}.\\
\texttt{doDSE[ilist, clist [, props, vertexTest, opts]]} & \texttt{vertexTest} is a function for determining if a vertex respects the symmetries of the Lagrangian. \texttt{props} is a list of allowed propagators given in the form \texttt{\{\{field1a, field1b\}, \{field2a, field2b\}, ...\}}. \texttt{doDSE} accepts the options \texttt{specificFieldDefinitions} and \texttt{sourcesZero} (prevents the replacement of super-field propagators and vertices when set to \texttt{False}).\\
\texttt{shortExpression[expr, opts]} \quad \texttt{sE[expr, opts]} & Rewrites a \textit{DoDSE} expression into a shorter form using \texttt{\$bareVertexSymbol}, \texttt{\$vertexSymbol} and \texttt{\$propagatorSymbol} for representation. Options of \texttt{Style} can be given.\\
\texttt{DSEPlot[expr, ilist~[,fRules,len,opts]]} & Plots graphs. \texttt{expr} is an expression containing \texttt{op} functions, \texttt{ilist} the list of interactions and \texttt{fRules} a list of options for plotting individual fields. \texttt{len} determines how many graphs are shown in one line. If \texttt{fRules} is not given, the lines are named according to the fields. Possible options are: \texttt{output->List}, to get the result in list form, and \texttt{indexStyle} and \texttt{factorStyle} to change the style of the indices and the prefactors (e.g. font size or color).
\ecommandlist

\bcommandlist
\multicolumn{2}{c}{\textbf{Functions for the individual computation steps}}\\
%&\textbf{Functions for the individual computation steps} \\
 \hline
  Command & Description\\
 \hline
  \texttt{generateAction[ilist[,flist]]} & Generates the action in internal representation from the interactions of the theory given in \texttt{ilist}. For mixed propagators \texttt{flist} specifies explicitly the type of fields in the form \texttt{\{boson1, boson2, ..., \{fermion1, antifermion1\}, \{fermion2, antifermion2\}, ...\}}.\\ 
  \texttt{deriv[expr,dlists]} & Differentiate \texttt{expr} with respect to the fields in \texttt{dlists}.\\
  \texttt{replaceFields[expr]} & Replaces the fields in \texttt{expr} by the corresponding expressions after 
the first differentiation is done to change from full to 1PI Green functions.\\
  \texttt{identifyGraphs[expr[, compareGraphs->cfunc]]} & Adds up equivalent graphs in \texttt{expr}. \texttt{cfunc} can be \texttt{compareGraphs} (standard) or \texttt{compareGraphs2}, the latter being necessary for mixed propagators but taking longer.\\
  \texttt{setSourcesZero[expr, flist [, props, vertexTest]]} & Sets the external fields in \texttt{flist} to zero, i.e. only physical propagators and vertices are left. \texttt{vertexTest} is a function for determining if a vertex respects the symmetries of the Lagrangian. \texttt{props} is a list of allowed propagators given in the form \texttt{\{\{field1a, field1b\}, \{field2a, field2b\}, ...\}}.\\
  \texttt{orderFermions[expr]} & Orders derivatives with respect to Grassmann fields such that the anti-fields are left of the fields thereby possibly giving a minus sign. \texttt{expr} is an \texttt{op}-function or a sum of those. Bare vertices are not affected by the ordering.
\ecommandlist

\bcommandlist
\multicolumn{2}{c}{\textbf{Functions for checks and tools}}\\
 \hline
  Command & Description\\
 \hline
  \texttt{countTerms[expr]} & Counts the number of terms appearing in the expression.\\
  \texttt{fieldQ[f]} & Determines if expression \texttt{f} is defined as a field.\\
  \texttt{bosonQ[f]} & Determines if expression \texttt{f} is defined as a bosonic field.\\
  \texttt{fermionQ[f]} & Determines if expression \texttt{f} is defined as a fermionic field.\\
  \texttt{antiFermionQ[f]} & Determines if expression \texttt{f} is defined as an anti-field to a fermionic field.\\
  \texttt{checkFields[expr]} & Checks if all fields in the expression are defined\\
  \texttt{checkIndices[expr]} & Checks if an index appears more often than twice.\\
  \texttt{checkSyntax[expr]} & Checks if \texttt{expr} has the correct syntax, i.e. \texttt{op} functions only contain propagators, vertices and fields.\\
  \texttt{checkAction[expr]} & Checks if all indices appear exactly twice, the syntax is ok and all fields are defined.\\
  \texttt{checkAll[expr]} & Performs a series of checks on \texttt{expr} (\texttt{checkIndices}, \texttt{checkSyntax}, \texttt{checkFields}).\\
  \texttt{defineFields[flist]} & Defines the fields of the action that are given in \texttt{flist} as single entries for bosons and grouped by braces for fermions.\\
     \texttt{\$vertexSymbol} & Symbol representing a vertex in \texttt{shortExpression}. Standard value: \texttt{$\Gamma$}.\\
   \texttt{\$bareVertexSymbol} & Symbol representing a bare vertex in \texttt{shortExpression}. Standard value: \texttt{S}.\\
   \texttt{\$PropagatorSymbol} & Symbol representing a propagator in \texttt{shortExpression}. Standard value: \texttt{$\Delta$}.
\ecommandlist

\bibliographystyle{utphys}
\bibliography{literature}

\providecommand{\href}[2]{#2}\begingroup\raggedright\begin{thebibliography}{10}

\bibitem{Dyson:1949ha}
F.~J. Dyson,
{\em Phys. Rev.} {\bf 75} (1949)  1736--1755.
%%CITATION = PHRVA,75,1736;%%.

\bibitem{Schwinger:1951ex}
J.~S. Schwinger,
{\em Proc. Nat. Acad. Sci.} {\bf 37} (1951)  452--455.
%%CITATION = PNASA,37,452;%%.

\bibitem{Alkofer:2008tt}
R.~Alkofer, C.~S. Fischer, F.~J. Llanes-Estrada, and K.~Schwenzer,
  \href{http://arxiv.org/abs/0804.3042}{{\tt 0804.3042 [hep-ph]}}.
Ann. Phys., in print.
%%CITATION = 0804.3042;%%.

\bibitem{Alkofer:2008et}
R.~Alkofer, C.~S. Fischer, and R.~Williams,
\href{http://arxiv.org/abs/0804.3478}{{\tt 0804.3478 [hep-ph]}}.
%%CITATION = 0804.3478;%%.

\bibitem{Watson:2006yq}
P.~Watson and H.~Reinhardt,
  \href{http://dx.doi.org/10.1103/PhysRevD.75.045021}{{\em Phys. Rev.} {\bf
  D75} (2007)  045021},
\href{http://arxiv.org/abs/hep-th/0612114}{{\tt hep-th/0612114}}.
%%CITATION = HEP-TH/0612114;%%.

\bibitem{VillalbaChavez:2008dv}
S.~Villalba-Chavez, R.~Alkofer, and K.~Schwenzer,
\href{http://arxiv.org/abs/0807.2146}{{\tt 0807.2146 [hep-th]}}.
%%CITATION = 0807.2146;%%.

\bibitem{Min:1985bx}
H.~Min, T.~Lee, and P.~Y. Pac,
\href{http://dx.doi.org/10.1103/PhysRevD.32.440}{{\em Phys. Rev.} {\bf D32}
  (1985)  440}.
%%CITATION = PHRVA,D32,440;%%.

\bibitem{Fazio:2001rm}
A.~R. Fazio, V.~E.~R. Lemes, M.~S. Sarandy, and S.~P. Sorella,
  \href{http://dx.doi.org/10.1103/PhysRevD.64.085003}{{\em Phys. Rev.} {\bf
  D64} (2001)  085003},
\href{http://arxiv.org/abs/hep-th/0105060}{{\tt hep-th/0105060}}.
%%CITATION = HEP-TH/0105060;%%.

\bibitem{Baulieu:1981sb}
L.~Baulieu and J.~Thierry-Mieg,
\href{http://dx.doi.org/10.1016/0550-3213(82)90454-0}{{\em Nucl. Phys.} {\bf
  B197} (1982)  477}.
%%CITATION = NUPHA,B197,477;%%.

\bibitem{ThierryMieg:1985yv}
J.~Thierry-Mieg,
\href{http://dx.doi.org/10.1016/0550-3213(85)90562-0}{{\em Nucl. Phys.} {\bf
  B261} (1985)  55}.
%%CITATION = NUPHA,B261,55;%%.

\bibitem{Mertig:1990an}
R.~Mertig, M.~Bohm, and A.~Denner,
\href{http://dx.doi.org/10.1016/0010-4655(91)90130-D}{{\em Comput. Phys.
  Commun.} {\bf 64} (1991)  345--359}.
%%CITATION = CPHCB,64,345;%%.

\bibitem{Hahn:2000kx}
T.~Hahn, \href{http://dx.doi.org/10.1016/S0010-4655(01)00290-9}{{\em Comput.
  Phys. Commun.} {\bf 140} (2001)  418--431},
\href{http://arxiv.org/abs/hep-ph/0012260}{{\tt hep-ph/0012260}}.
%%CITATION = HEP-PH/0012260;%%.

\bibitem{Caravaglios:1995cd}
F.~Caravaglios and M.~Moretti,
  \href{http://dx.doi.org/10.1016/0370-2693(95)00971-M}{{\em Phys. Lett.} {\bf
  B358} (1995)  332--338},
\href{http://arxiv.org/abs/hep-ph/9507237}{{\tt hep-ph/9507237}}.
%%CITATION = HEP-PH/9507237;%%.

\bibitem{Kanaki:2000ey}
A.~Kanaki and C.~G. Papadopoulos,
  \href{http://dx.doi.org/10.1016/S0010-4655(00)00151-X}{{\em Comput. Phys.
  Commun.} {\bf 132} (2000)  306--315},
\href{http://arxiv.org/abs/hep-ph/0002082}{{\tt hep-ph/0002082}}.
%%CITATION = HEP-PH/0002082;%%.

\bibitem{Mangano:2002ea}
M.~L. Mangano, M.~Moretti, F.~Piccinini, R.~Pittau, and A.~D. Polosa, {\em
  JHEP} {\bf 07} (2003)  001,
\href{http://arxiv.org/abs/hep-ph/0206293}{{\tt hep-ph/0206293}}.
%%CITATION = HEP-PH/0206293;%%.

\bibitem{Pawlowski:2005xe}
J.~M. Pawlowski, \href{http://dx.doi.org/10.1016/j.aop.2007.01.007}{{\em Annals
  Phys.} {\bf 322} (2007)  2831--2915},
\href{http://arxiv.org/abs/hep-th/0512261}{{\tt hep-th/0512261}}.
%%CITATION = HEP-TH/0512261;%%.

\bibitem{Itzykson:1980ft}
C.~Itzykson and J.-B. Zuber, {\em Quantum Field Theory}.
\newblock Dover Publications, Mineola, New York, 1980.

\bibitem{Roberts:1994dr}
C.~D. Roberts and A.~G. Williams, {\em Prog. Part. Nucl. Phys.} {\bf 33} (1994)
   477--575,
\href{http://arxiv.org/abs/hep-ph/9403224}{{\tt hep-ph/9403224}}.
%%CITATION = HEP-PH 9403224;%%.

\bibitem{Alkofer:2000wg}
R.~Alkofer and L.~von Smekal, {\em Phys. Rept.} {\bf 353} (2001)  281,
\href{http://arxiv.org/abs/hep-ph/0007355}{{\tt hep-ph/0007355}}.
%%CITATION = HEP-PH 0007355;%%.

\bibitem{Rivers:1988pi}
R.~J. Rivers, {\em Path Integrals Methods in Quantum Field Theory}.
\newblock Cambridge University Press, Cambridge, 1988.

\bibitem{Schleifenbaum:2004id}
W.~Schleifenbaum, A.~Maas, J.~Wambach, and R.~Alkofer, {\em Phys. Rev.} {\bf
  D72} (2005)  014017,
\href{http://arxiv.org/abs/hep-ph/0411052}{{\tt hep-ph/0411052}}.
%%CITATION = HEP-PH 0411052;%%.

\bibitem{Wolfram:1999}
S.~Wolfram, {\em The Mathematica Book}.
\newblock Cambridge University Press, 1999.

\bibitem{Alkofer:2004it}
R.~Alkofer, C.~S. Fischer, and F.~J. Llanes-Estrada, {\em Phys. Lett.} {\bf
  B611} (2005)  279--288,
\href{http://arxiv.org/abs/hep-th/0412330}{{\tt hep-th/0412330}}.
%%CITATION = HEP-TH 0412330;%%.

\bibitem{Alkofer:2008jy}
R.~Alkofer, M.~Q. Huber, and K.~Schwenzer,
\href{http://arxiv.org/abs/0801.2762}{{\tt 0801.2762 [hep-th]}}.
%%CITATION = 0801.2762;%%.

\end{thebibliography}\endgroup

\end{document}